\documentclass[11pt]{article}

\makeatletter
\def\eqnarray{\stepcounter{equation}\let\@currentlabel=\theequation
\global\@eqnswtrue
\global\@eqcnt\z@\tabskip\@centering\let\\=\@eqncr
$$\halign to \displaywidth\bgroup\@eqnsel\hskip\@centering
  $\displaystyle\tabskip\z@{##}$&\global\@eqcnt\@ne
  \hfil${\;##\;}$\hfil
  &\global\@eqcnt\tw@ $\displaystyle\tabskip\z@{##}$\hfil
   \tabskip\@centering&\llap{##}\tabskip\z@\cr}
\makeatother

\newcommand{\smfrac}[2]{{\textstyle{#1\over#2}}}
\def\half{\smfrac{1}{2}}
\def\curl{\widetilde{{\rm curl}}}

\newcommand{\ii}{{\rm i}}

\begin{document}
%
\title{Local freedom in the gravitational field revisited}
\author{ Mar\'{\i }a Jes\'us Pareja\thanks{
     Institute for Astronomy and Astrophysics,
     University of T\"ubingen,
     Auf der Morgenstelle 10, D-72076 T\"ubingen, Germany.
     Email: pareja@tat.physik.uni-tuebingen.de}~~and
Malcolm A.H. MacCallum\thanks{School of Mathematical Sciences,
     Queen Mary, University of London,
     Mile End Road,
     London E1 4NS,
     U.K.
Email: m.a.h.maccallum@qmul.ac.uk}}
\date{}
\maketitle
\begin{abstract}
  Maartens {\it et al.\/}\@ \cite{MES} gave a covariant
  characterization, in a 1+3 formalism based on a perfect fluid's
  velocity, of the parts of the first derivatives of the curvature
  tensor in general relativity which are ``locally free'', i.e.\ not
  pointwise determined by the fluid energy momentum and its
  derivative.  The full decomposition of independent curvature
  derivative components given in earlier work on the spinor approach
  to the equivalence problem enables analogous general results to be
  stated for any order: the independent matter terms can also be
  characterized. Explicit relations between the two sets of results
  are obtained. The $24$ Maartens {\it et al.}  locally free data are
  shown to correspond to the $\nabla \Psi$ quantities in the spinor
  approach, and the fluid terms are similarly related to the
  remaining $16$ independent quantities in the first derivatives of
  the curvature.
\end{abstract} 
 
\section{Introduction and discussion of the general case}

Maartens {\it et al.\/}\@ \cite{MES} defined ``locally free'' terms in
the curvature tensor and its first derivative, in general relativity,
to be those not pointwise algebraically determined by the matter
fields and their first derivatives. This definition can be extended in
an obvious way to any order of differentiation. It is
distinct from consideration of the free data in initial value problems
for the Einstein equations, and from functional independence over the
manifold or its frame bundle. Note that the quantities found at order
$n$ will in general be subject to differential equations
(integrability conditions) at order $(n+1)$.

This definition can be related to the decomposition of the curvature
tensor and its derivatives used in classifying metrics and testing
their equivalence which is described, for example, in \cite{SKMHH},
chapter 9.  The locally free quantities and the matter terms together
contain all the invariant classificatory information.  Testing local
isometry of two given spacetimes can be reduced to checking the
consistency of a set of equations between the two Riemann tensors and
their covariant derivatives, which for efficiency is best done using a
canonically chosen frame.

For the curvature and first derivative in the perfect fluid case,
Maartens {\it et al.\/}\@ gave a covariant characterization, in a 3+1
formalism based on the invariantly-defined fluid velocity $u^a$, of
the locally free quantities. (Note that full classification might need
higher derivatives: for general energy-momenta, an example requiring
the 5th derivative is known \cite{Skea}.) Their work was done with
cosmology in mind. The electric and magnetic parts of the Weyl tensor
defined relative to $u^a$, $E_{ab}$ and $H_{ab}$, covariantly
represent the locally free parts of the curvature.

Performing a complete covariant $1+3$ decomposition of $\nabla_c
E_{ab}$ and $\nabla_c H_{ab}$, where $\nabla_c$ denotes the covariant
derivative in the $c$ direction, into time derivatives, spatial curls,
divergences and ``distortions'' (full definitions appear in the next
section), it was shown that the locally free parts of the derivative
of the curvature can be covariantly represented by the distortions and
the spatial curls of $E_{ab}$ and $H_{ab}$. These provide $24$ free
data.  The relations that constrain the remaining first derivatives
are the Ricci and Bianchi identities, which imply that the remaining
parts of the derivatives of $E_{ab}$ and $H_{ab}$ are not locally
free.

MacCallum and {\AA}man \cite{MA} showed, using a two-component spinor
formalism, that in any spacetime a minimal set of components of the
derivatives of the Riemann curvature of a given order $m$ can be
specified such that all components of the derivatives of the Riemann
curvature of order $n$ can be expressed algebraically in terms of
these sets for $m \leq n$.  This generalized a result by Penrose
\cite{P} to the non-vacuum case.  They also gave a specific
prescription for these sets. The chosen quantities are formed by
totally symmetrized spinor derivatives and contractions. Terms not in
the minimal sets can be calculated using the Ricci and Bianchi
identities and consequent higher order integrability conditions, which
are the only equations relevant in the most general case.
 The Bianchi
identities take the form
\begin{eqnarray}
\nabla^D_{\ C'} \Psi_{ABCD} & = & \nabla^{D'}_{\ (C} \Phi_{AB)C'D'}\ ,
\label{spB1}\\
\nabla^{BB'} \Phi_{ABA'B'} & = & -3 \nabla_{AA'} \Lambda 
\label{spB2} \ .
\end{eqnarray}

The chosen minimal set for the components of the Riemann tensor and
its derivatives up to the $n$-th, which we denote collectively by
$\nabla^n R$, can be described as follows \cite{MA}, using the
Newman--Penrose quantities $\Psi_{ABCD}, \Phi_{ABA'B'}$ and $\Lambda$,
and defining a `totally symmetrized' spinor to be one symmetrized over
all its free dashed indices and also over all its free undashed
indices.  We take the following components for $n \geq q \geq 0$:
\begin{list}{}{\setlength{\itemindent}{0pt}%
\setlength{\itemsep}{0pt}\setlength{\parsep}{0pt}}
\item[(i)]{The totally symmetrized $q$th derivatives of $\Psi_{ABCD}$}.
\item[(ii)]{The totally symmetrized $q$th derivatives of $\Phi_{ABA'B'}$}.
\item[(iii)]{The totally symmetrized $q$th derivatives of $\Lambda$}.
\item[(iv)]{For $q \geq 1$, the totally symmetrized ($q-1$)th
derivatives of
\begin{equation}
\label{Xidef}
\Xi_{ABCD'} \equiv \nabla^{D'}_{\ (C} \Phi_{AB)C'D'}
(=\nabla^D\null_{D'}\Psi_{DABC})
\end{equation}
which is one side of (\ref{spB1}).}
\item[(v)]{For $q \geq 2$, the d'Alembertian
    $\nabla^{AA'}\nabla_{AA'}$ applied to all the quantities
    calculated for the derivatives of order $q-2$.}
\end{list}

It is worth recalling that the symmetrization removes all trace terms.
The number of $n$-th derivative terms in this set in a general
spacetime is $(n + 1)(n + 4)(n + 5),$ which is less, and for large $n$
much less, than the total number of components of the $n$th
derivatives of the independent components of the Riemann tensor, $20$
x $4^n$. Particular cases may satisfy some restrictions on the algebraic
structure of the Weyl and Ricci tensors and/or differential
identities which reduce the number of free quantities.

It is immediately obvious that the locally free data up to order $n$
are given by the terms of type (i) and those of type (v) obtained
from them, for all forms of matter and all values of $n$.

In the remainder of this paper we examine the relationship between the
above set for $n=1$ and the results of Maartens {\it et
  al.\/} The separation of locally free and matter terms is not
unique, since linear independence is not affected by addition of other
terms to the independent quantities. The two methods described above
turn out to lead to different linear combinations of terms, but, as
one would expect, the locally free data sets agree modulo terms which
are not locally free.

Following \cite{MES}, we restrict ourselves to the case where the
matter content is a perfect fluid, which is of special importance in
an astrophysical context. This reduces the number of independent
components in the matter tensor from 10 to 5 (the density, pressure
and three of the components of the 4-velocity), and the number of first
derivatives correspondingly from 40 to 20, in both cases subject to
the 4 contracted Bianchi identities, leaving 36 and 16 free components
respectively. Choosing an orthonormal tetrad such that its timelike
basis vector is aligned with the 4-velocity $u^a$, the number of
independent non-zero components in the energy-momentum reduces to 2.
The Weyl tensor has 10 independent components: 16 of their derivatives
are fixed by the Bianchi identities and the remaining 24 are locally
free (and correspond to the 24 terms of type (i) above for $n=1$).

To make the correspondence one needs to represent $u^a$ in the spinor
formalism. The obvious way, with the usual correspondence between the
spinor basis and a null basis (\cite{SKMHH}, \S\ 3.6), is to choose
the null tetrad so that the real null basis vectors are coplanar with
$u^a$; this implies the choice of two basis vectors, rather than the
one fixed in the $1+3$ form.  To reflect this choice in the $1+3$
system then requires a choice of a spacelike vector. For simplicity we
deal with this by embedding the $1+3$ approach in an orthonormal
tetrad formalism, and relating that explicitly to the null tetrad as
in \cite{SKMHH}, (3.12).  The proof of the correspondence between the
two sets of results described above is then a direct calculation on
components. The relations are nevertheless tensorial in character: the
choice of bases is no more restrictive than proving tensor equations
by checking them in a specific coordinate basis. Note that we have not
made an invariant choice of the spatial vectors, so this method does
not correspond to carrying out the procedure for choosing a frame in
the equivalence problem.

In section 2 we give the results of \cite{MES} in more detail, and in
section 3 those of \cite{MA}. Section 4 links the two as outlined
above.  Throughout we follow the notation and conventions of
\cite{SKMHH}, unless otherwise stated.  We find as expected that the
Maartens {\it et al.\/}\@ \cite{MES} locally free data correspond with
the 24 free components of type (i) above.  The 16 remaining parts of
the first derivative of the Riemann tensor $\nabla R$, similarly
correspond with the 16 algebraically independent quantities which
occur in (ii) to (iv) in the perfect fluid case.

If the $1+3$ decomposition via Young tableaux and removal of trace
terms, as described in \S\ 3 of \cite{MES}, were carried out for $n>1$,
the results could similarly, if laboriously, be compared with the
prescription (i)-(v) above.

\section{Covariant $1+3$ characterization of the {\it locally free}
parts of the Riemann tensor and its covariant derivative}

The energy-momentum tensor for perfect fluid is $T_{i j} = (\mu + p)
u_{i} u_{ j} + p g_{i j}\ ,$ where $p$ denotes the pressure and $\mu$
the energy density. We shall assume that $\mu+p
\neq 0$, so that the 4-velocity $u^a$ is well and covariantly defined.
(In the case $\mu+p = 0$ where the 4-velocity is indeterminate, the
matter being just a cosmological constant, the kinematic quantities
below are also indeterminate.)

Projecting the covariant derivatives of $u^a$ parallel and orthogonal
to it, using $h_{ab} = g_{ab} + u_a u_b$, we obtain the spatial
``kinematic quantities'' $\Theta$ (expansion), $\dot{u}_a$
(4-acceleration), $\omega_{ab}$ (vorticity) and $\sigma_{ab}$ (shear):
\begin{eqnarray}
u_{a;b} & = & \omega_{ab}+\sigma_{ab}+\smfrac{1}{3}\Theta h_{ab}-\dot{u}_au_b,
\nonumber \\ 
\dot{u}_a &:= & u_{a;b}u^b=Du_a/d\tau, \quad \dot{u}_au^a=0, \quad
\omega_{ab}  := u_{[a;b]}+\dot{u}_{[a}u_{b]},\quad
 \omega_{ab}u^b=0, \nonumber\\
\sigma_{ab} & := & u_{(a;b)}+\dot{u}_{(a}u_{b)}-\smfrac{1}{3}\Theta
 h_{ab},\quad \sigma_{ab}u^b=0, \quad \Theta := u^{a}{}_{;a} . \nonumber
\end{eqnarray}
Using the covariant 3-dimensional permutation tensor $\eta_{abc}
\equiv \eta_{abcd} u^d$, we define the vorticity vector $\omega_a
\equiv \eta_{abc} \omega^{bc}$.

Using the field equations, the trace-less Ricci tensor
$S_{ab}=R_{ab}-\smfrac{1}{4}Rg_{ab}$ and the curvature scalar are
related to the perfect fluid quantities by
$$S_{44} = \smfrac{3}{ 4} (\mu + p), \quad
S_{\alpha \beta} =\smfrac{1}{ 4} (\mu + p) h_{\alpha \beta}, \quad
\mbox{and} \quad R = \mu - 3 p\ ,$$
where components in the $u^a$
direction are labelled 4 and those orthogonal to $u^a$ by Greek
indices, $\alpha, \beta \in \{1,2,3\}$. These parts of the curvature
are not locally free; the locally free parts are the electric and
magnetic parts of the Weyl tensor, $E_{ab} \equiv C_{acbd} u^c u^d$
and $H_{ab} \equiv \half \eta_{acd} C^{cd}_{\ \ be} u^e$, which
satisfy $E_{ab} = E_{\langle ab \rangle}$ and $H_{ab} = H_{\langle ab
  \rangle}$, where angle brackets $_{\langle ... \rangle}$ denote the
totally symmetrized, trace-free and spatially projected part, e.g.\ 
$T_{\langle ab \rangle} \equiv h_{(a}^{\ c} h_{b)}^{\ d} T_{cd} -
\smfrac{1}{ 3} h_{cd} T^{cd} h_{ab}$, and round brackets denote
symmetrization.

We denote the partial derivative by a comma, and covariant derivative
by a semicolon. $\widetilde{\nabla}$ will denote the projected
covariant derivative operator $\nabla$, in which all indices of the
resulting quantity are projected with $h_{ab}$.  The tilde is to
remind us that this derivative acts in a spatial section only when
the vorticity vanishes, while the $\nabla$ reminds us of the
usual notation in three-dimensional vector calculus. Then the
derivative of any traceless symmetric tensor
$B_{ab}$ has the following covariant decomposition \cite{MES}:
\begin{equation}
B_{ab;c} = - \dot{B}_{\langle ab \rangle} u_c - 2 u_{(a} B_{b)d} 
\dot{u}^d u_c + 2 u_{(a} \{ B_{b)}^{\ \ d} (\sigma_{dc} + \omega_{dc}) 
+ \smfrac{1}{ 3} \Theta B_{b)c} \} + \widetilde{\nabla}_c B_{ab}\ .
\end{equation}
\noindent Defining the divergence, curl and distortion of $B_{ab}$
respectively by
\[
(\widetilde{{\rm div}}\ B )_a \equiv \widetilde{\nabla}^b B_{ab} \ , \quad 
(\widetilde{{\rm curl}}\ B )_{ab} \equiv
 \eta_{cd(a} {\widetilde{\nabla}^c B_{b)}}^d \ , \quad
\widetilde{\nabla}_{\langle c} B_{ab \rangle}\ ,
\]
we have
\begin{equation}
\widetilde{\nabla}_c B_{ab} \equiv 
\widetilde{\nabla}_{\langle c} B_{ab \rangle} 
+ \smfrac{3}{ 5} ( \widetilde{{\rm div}}~ B )_{\langle a} h_{b \rangle c} 
-\smfrac{2}{ 3} ( \widetilde{{\rm curl}}~ B )_{d(a} \eta_{b)c}^{\ \ \ d}.
\end{equation}

The Bianchi identities (with perfect fluid), covariantly split, are:
\begin{eqnarray}
\dot{\mu} + (\mu + p) \Theta &=& 0 
  \label{B1}, \\
\widetilde{\nabla}_a p + (\mu + p) \dot{u}_a  &=& 0
 \label{B2}, \\
\dot{E}_{\langle ab \rangle} + \Theta E_{ab} 
  - (\widetilde{{\rm curl}}\ H)_{ab} - 3 \sigma_{c\langle a}
  {E_{b\rangle}}^{c}  &+& \omega_{d\langle a} {E_{b \rangle}}^{d}
  -2 \dot{u}^c \eta_{cd(a} {H_{b)}}^{d}\nonumber \\
     &=& - \smfrac{1}{2}(\mu + p)\sigma_{ab}
 \label{B3}, \\
\hspace{-1em}\dot{H}_{\langle ab \rangle} + \Theta H_{ab} + 
               (\widetilde{{\rm curl}}\ E)_{ab} - 3\
               \sigma_{c\langle a} {H_{b \rangle}}^{c}
 &+&\omega_{d\langle a} {H_{b \rangle}}^{d}
            + 2 \dot{u}^c \eta_{cd(a} {E_{b)}}^{d}=0 
%
  \label{B4},\\
( \widetilde{{\rm div}}\ E )_a - \eta_{abc} \sigma^b_{\ d} H^{cd} &+& 3
 H_a{}^b \omega_b = \smfrac{1}{ 3} \widetilde{\nabla}_a \mu 
  \label{B5},\\
\label{eq:WMaxwell3}
( \widetilde{{\rm div}}\ H )_a + \eta_{abc} \sigma^b_{\ d} E^{cd} &-& 3
 E_a{}^b \omega_b = (\mu + p)\omega_a
  \label{B6},
\end{eqnarray}
\noindent where the dot indicates the covariant derivative along the
  fluid $4$-velocity.

The covariant derivatives of the matter terms are
$$S_{44;c} = \smfrac{3}{ 4} (\mu + p)_{,c}\ ,\quad
  S_{\alpha\beta;c} = \smfrac{1}{ 4} (\mu + p)_{,c} h_{\alpha\beta}\ ,$$
$$S_{\alpha 4;4} = - (\mu + p) \dot{u}_\alpha\ ,\quad 
  S_{\alpha 4;\beta} = -(\mu + p) (\omega_{\alpha\beta} + \sigma_{\alpha\beta} 
   + \smfrac{1}{3} \Theta g_{\alpha\beta})\ ,$$
$$R_{;c} = (\mu - 3 p)_{,c}\ .$$

\noindent  Thus, the parts of $\nabla R$
determined pointwise by the matter (via the Bianchi identities) can be
covariantly $(1 + 3)$ characterized by $16$ fluid variables. These can
be chosen from the $8$ derivatives of $\mu$ and $p$, the $5$
components of $\sigma_{ab}$, the $3$ each of $\omega_a$ and
$\dot{u}_c$, and $\Theta$, by eliminating 4 determined from the others
by the contracted Bianchi identities (\ref{B1}) and (\ref{B2}). We may
also note that all the kinematic quantities are covariantly defined by
ratios of matter derivatives to $\mu + p$, which is itself defined by
the Ricci tensor.  If the spatial axes were determined, e.g.\ as
eigenvectors of $\sigma_{ab}$, the remaining components of the
kinematic quantities are then determined by Cartan invariants in the
sense of \cite{SKMHH}. These details are not discussed in \cite{MES}.

We observe that the divergences of the locally free fields $E_{ab}$
and $H_{ab}$ are determined pointwise by the matter terms and the
locally free fields via the Bianchi constraint identities (\ref{B5})
and (\ref{B6}).  The curls of the locally free fields, together with the
matter terms and the fields themselves, determine the time
derivatives of $E_{ab}$ and $H_{ab}$ via the Bianchi evolution
identities (\ref{B3}) and (\ref{B4}).  The distortions
$\widetilde{\nabla}_{\langle c} E_{ab \rangle}$ and
$\widetilde{\nabla} _{\langle c} H_{ab \rangle}$ are exactly the
parts of $\nabla_c E_{ab}$ and $\nabla_c H_{ab}$ which are missing
from the (Ricci and) Bianchi identities, and do not locally depend
algebraically on the matter fields.  Therefore the distortions and
curls of the electric and magnetic parts of the Weyl tensor provide a
covariant characterization of the locally free parts of the derivative
of the curvature tensor.

Since the distortions contain $2$ x $7 = 14$ independent components
and the curls contain $2$ x $5 = 10$ independent components, there are
at each point of the spacetime, in general, $24$ components of
$\nabla_e C_{abcd}$ and thus of $\nabla_e R_{abcd}$ {\it not constrained}
algebraically by the matter terms.

\section{Characterization of the Riemann spinor and its derivative}

At $q=0$ the split given earlier just tells us that the Riemann tensor
splits into a locally free part $\Psi_{ABCD}$ and matter parts
$\Phi_{ABA'B'}$ and $\Lambda$. The minimal set of
spinor first derivatives is:

\noindent (i) the totally symmetrized spinor derivative of the Weyl spinor
$$\nabla_{(X}^{\ \ \ X'} \Psi_{ABCD)}\ ,$$
(ii) the totally symmetrized spinor derivative of the Ricci spinor
$$\nabla_{(X}^{\ \ (X'} \Phi_{AB)}^{\phantom{AB)}A'B')}\ ,$$
(iii) the spinor derivative of the scalar curvature
$$\nabla_{X}^{\ \ X'} \Lambda\ ,$$
(iv) the `curl' $\Xi_{CABC'}$ of the Ricci spinor defined 
as in (\ref{Xidef}).

This minimal set has $60$ independent quantities, whereas the
total number of components is $80.$ The specialization of (ii)-(iv)
for a perfect fluid is deferred to the next section.

Hereafter we follow the convention that components of totally
symmetrized spinors are labelled by one primed and one unprimed index
whose numerical values count those spinor indices which are projected
onto the spinor basis element $\iota$ or its conjugate, so that, for
example, we write

$$\nabla \Psi_{20'} \equiv \nabla_{(A|X'} \Psi_{|BCDE)} o^A o^B o^C
\iota^D \iota^E \bar{o}^{X'}.$$

\section{Correspondence between the covariant
  $1 + 3$ and spinor decompositions of $\nabla R$.}

We denote the orthonormal tetrad\footnote{The notation here is changed
  from that of \cite{SKMHH} to avoid confusion between components of
  the tetrad and of the electric part of the Weyl tensor} by
$\{{e_{a}}^{i}\}$ where later Latin indices $i,j$ now refer to
coordinate components, and its dual covector basis by
$\{{\omega^{a}}_{i}\}$.
The timelike basis vector is chosen
to be the fluid $4$-velocity, so $u^{i} = {e_{4}}^{i}$
and $u_{i} = -{\omega^{4}}_{ i} .$
In the spinor (null tetrad) approach we  use a null tetrad
$\{m^{i},\bar{m}^{i}, l^{i},k^{i}\}$
such that the fluid $4$-velocity is 
${e_{4}}^{i} = u^{i} = {1 \over \sqrt{2}} (k^{i} + l^{i}).$

Accordingly, we consider the ONT - null tetrad transformation\\
\begin{minipage}[l]{.40\linewidth}
\begin{eqnarray}
m^{i} & = & {1 \over \sqrt{2}} ({e_{1}}^{i} - \ii\ {e_{2}}^{i})\nonumber\\
\bar{m}^{i} & = & {1 \over \sqrt{2}} ({e_{1}}^{i}
                   + \ii\ {e_{2}}^{i})\nonumber\\
l^{i} & = & {1 \over \sqrt{2}} ({e_{4}}^{i} - {e_{3}}^{i})\nonumber\\
k^{i} & = & {1 \over \sqrt{2}} ({e_{4}}^{i} + {e_{3}}^{i})\nonumber
\end{eqnarray}
\end{minipage}\hfill
\begin{minipage}[b]{.16\linewidth}
\mbox{and its dual}
\end{minipage}
\begin{minipage}[r]{.42\linewidth}
\begin{eqnarray}
\bar{m}_{i} & = & {1 \over \sqrt{2}} ({\omega^{1}}_{i} + \ii 
\ {\omega^{2}}_{i})\nonumber\\
{m}_{i} & = & {1 \over \sqrt{2}} ({\omega^{1}}_{i} - 
\ii\ {\omega^{2}}_{i})\nonumber\\[-4.5pt]
\label{ONT2NP}\\[-5.5pt]
-k_{i} & = & {1 \over \sqrt{2}} ({\omega^{4}}_{i} -{\omega^{3}}_{i})\nonumber\\
-l_{i} & = & {1 \over \sqrt{2}} ({\omega^{4}}_{i} +
{\omega^{3}}_{i})\ ,\nonumber
\end{eqnarray}
\end{minipage}\\

\noindent so that the metric written in terms of the null tetrad is
$$g_{ij} = 2 m_{(i} \bar{m}_{ j)} - 2 k_{(i} 
l_{ j)}\ .$$

When we restrict ourselves to the special case of perfect fluid source
this choice will lead to restrictions on the algebraic structure of
the Ricci spinor, and consequent differential identities. One obtains
the following relations among the components of the Ricci spinor
\begin{eqnarray}
\Phi_{11'} & = & \half \Phi_{00'}\nonumber, \quad \Phi_{22'} = \Phi_{00'},
\quad
\Phi_{ab'} = 0\ \ \ \forall a \neq b \ \ (a, b \in \{0,1,2\}).\label{spmatter}
\end{eqnarray}
There is only one independent component in $\Phi_{ABA'B'}$,
whose relation with the fluid pressure and energy density is
$\Phi_{00'} = \smfrac{1}{4} (\mu + p)$. There is also the curvature
scalar $\Lambda = \smfrac{1}{ 24} R = \smfrac{1}{ 24} (\mu - 3 p)$.

To complete the correspondence of the ($1+3$) and spinor decompositions of
the curvature part of $\nabla R$, we define\footnote{Changing the sign of
  $Q$ as compared with \cite{SKMHH}.}
\begin{equation}
\label{eq:3.62}Q_{ac}\equiv C_{abcd}^{*}u^bu^d = E_{ac}+\ii %
H_{ac},
\end{equation}
where $C_{abcd}^{*}$ is the self-dual part of the Weyl tensor. Then we have
\begin{eqnarray}
C^{\ast}_{abcd}=8u_{[a}Q_{b][d}u_{c]}+2g_{a[c}Q_{d]b}-2g_{b[c}Q_{d]a}\quad
 \nonumber \\
+2\ii\varepsilon_{abef}u^eu_{[c}Q_{d]}{}^f
 +2\ii\varepsilon_{cdef}u^eu_{[a}Q_{b]}{}^f.
\label{eq:3.63}
\end{eqnarray}
The spinor equivalent of $C_{abcd}^{*}$ is $\varepsilon_{A'B'}
\varepsilon_{C'D'}\Psi_{ABCD}$. Contracting with 
basis vectors we obtain
\begin{eqnarray}
\hspace*{-1em}\Psi_0 &=& (Q_{11}-Q_{22}-2\ii Q_{12})/2, \quad \Psi_1=(\ii
Q_{23}-Q_{13})/2, \quad \Psi_2=Q_{33}/2, \nonumber \\
\label{Psi2Q}
\hspace*{-1em}\Psi_4 &=&  (Q_{11}-Q_{22}+2\ii Q_{12})/2,
\quad\Psi_3=(\ii Q_{23}+Q_{13})/2.
\end{eqnarray}

\subsection{Correspondence between the ``locally free'' parts of the
  first derivatives}

The locally free parts in spinor form as chosen above, the terms
$\nabla \Psi_{ab'}$, are linear combinations of distorsions and curls
of $E_{ab}$ and $H_{ab}$ (i.e.\ of $Q_{ab}$), which are the independent
locally free parts in the $1+3$ form as chosen by \cite{MES}, together
with time derivatives of $E_{ab}$ and $H_{ab}$ and products of
kinematic quantities with $E_{ab}$ and $H_{ab}$.  The time derivatives
are determined from matter terms and curls of $E_{ab}$ and $H_{ab}$
via the Bianchi evolution identities (\ref{B3}) and (\ref{B4}), and
the kinematic quantities are matter terms, so these parts are not
locally free. The divergences of $E_{ab}$ and $H_{ab}$ do not appear
in the $\nabla \Psi$.

We can obtain these results, which are independent of the perfect
fluid assumption, as follows.  Differentiating the spinor
form of $C_{abcd}^{*}$ gives just derivatives of $\Psi$, whereas
differentiating (\ref{eq:3.63}) gives combinations of $\nabla_c
E_{ab}$, $\nabla_c H_{ab}$ and products of kinematic quantities with
$E_{ab}$ and $H_{ab}$. That the 12 locally free complex terms $\nabla
\Psi_{ab'}$ of type (i) above give the 24 real ($1+3$) locally free
quantities, modulo terms which are not locally free, and vice versa,
can then be shown explicitly.

To do this, we note first that the $\Psi_{ab'}$ are rather simple
combinations of the $\Psi_{ABCD;EE'}$. For example, using ${}_{,m}$
and ${}_{;m}$ for partial and covariant derivative, respectively, in
the ${\bf m}$ direction, etc, we find
$\nabla\Psi_{21'}=(3\Psi_{0011;m}+2\Psi_{0001;\ell})/5$. We write the
($1+3$) quantities in terms of $Q_{ab}$: real and imaginary parts can
readily be separated as a final step. As the independent components of
$Q_{\langle abc \rangle}$ we take the terms $Q_{\langle 111 \rangle}$,
$Q_{\langle 112 \rangle}$, $Q_{\langle 113 \rangle}$, $Q_{\langle 123
  \rangle}$, $Q_{\langle 122 \rangle}$, $Q_{\langle 222 \rangle}$, and
$Q_{\langle 223 \rangle}$, and as the independent components of
$\curl\ Q$ the terms $\curl\ Q_{12}$, $\curl\ Q_{23}$, $\curl\ 
Q_{31}$, $\curl\ Q_{11}$, and $\curl\ Q_{22}$.  Using (\ref{Psi2Q}),
we then obtain
\begin{eqnarray}
\label{Psi00}
{2\sqrt{2}} \nabla \Psi_{00'} & = & 
(\widetilde {\nabla}_{\langle 3} Q_{11\rangle} - \widetilde{\nabla}_
{\langle 3} Q_{22\rangle}-2\ii \widetilde{\nabla}_{\langle 3} Q_{12\rangle})
+ (\dot{Q}_{11} -
\dot{Q}_{22} -2\ii \dot{Q}_{12}) \nonumber \\
& & 
 + \smfrac{4}{3} (\widetilde{{\rm curl}}~ Q)_{12}
 - \smfrac{2\ii}{3} ((\widetilde{{\rm curl}}~ Q)_{11}
 - (\widetilde{{\rm curl}}~Q)_{22})
 \nonumber \\
& &  + \ \mbox{matter\ terms} \times 
(E_{ab}, H_{ab}),
\end{eqnarray}
and 11 other complex equations. Displaying these in full would be
lengthy without being illuminating, so we omit the rest.

We now have to show the linear independence claimed. We give an
abbreviated description of the results.  The symbol $\cong$ will mean
equality modulo $\dot{Q}_{ab}$ and products of kinematic quantities
with $Q_{ab}$, which are not considered locally free derivative terms
in either approach.
  
On taking the combinations $\nabla \Psi_{k0'} \pm \nabla
\Psi_{(5-k)1'}, ~k=0,\,1,\ldots,\,5$, the expressions for $\nabla
\Psi$ separate into 4 groups of 3 equations, two from each of two sets
of 6 components, which can be solved for the $1+3$ locally free
quantities. Using the notation $(30'-21')$ for the difference of
$\nabla \Psi_{30'}$ and $\nabla \Psi_{21'}$, and so on, we find that:
$(00'-51')$, $(11'-40')$ and $(20'-31')$ give $Q_{\langle 113
  \rangle}$, $Q_{\langle 223 \rangle}$, and $\curl\ Q_{12}$;
$(00'+51')$, $(11'+40')$ and $(20'+31')$ give $Q_{\langle 123
  \rangle}$, $\curl\ Q_{11}$ and $\curl\ Q_{22}$; $(21'-30')$,
$(10'+41')$ and $(01'+50')$ give $Q_{\langle 111 \rangle}$,
$Q_{\langle 122 \rangle}$ and $\curl\ Q_{23}$; and $(21'+30')$,
$(10'-41')$ and $(01'-50')$ give $Q_{\langle 112 \rangle}$,
$Q_{\langle 222 \rangle}$ and $\curl\ Q_{31}$; and conversely.

As an example we give the structure for one of these groups of
equations. The equations $(21'-30')$, $(10'+41')$ and $(01'+50')$ are
respectively
\begin{eqnarray}
5\sqrt{2}(\nabla \Psi_{21'} - \nabla \Psi_{30'}) &\cong&
-(5Q_{\langle 111
    \rangle} + 5Q_{\langle 122 \rangle}+\smfrac{4}{3}\curl\ Q_{23}),\\
5\sqrt{2}(\nabla \Psi_{10'} - \nabla \Psi_{41'}) &\cong&
5Q_{\langle 111
    \rangle} + 5Q_{\langle 122 \rangle}-\smfrac{8}{3}\curl\
  Q_{23},\\
\sqrt{2}(\nabla \Psi_{01'} +\nabla \Psi_{50'})&\cong&Q_ {\langle 111
    \rangle}- 3Q_{\langle 122 \rangle}.
\end{eqnarray}
For this triple the ($1+3$) derivatives are easily seen to be
independent combinations of the left sides of the equations. The same
is true for the other three sets.

Equation (\ref{Psi00}) gives one of the locally free spinor
derivatives in terms of ($1+3$) locally free quantities (two real
equations). We end this section with examples from the converse
set. We find
$$\widetilde {\nabla}_{\langle 2} Q_{11\rangle} \cong 
\frac{2\sqrt{2}\ii}{ 3} 
(\nabla \Psi_{10'} +3 \nabla \Psi_{01'} -2 \nabla \Psi_{21'} -2
\nabla \Psi_{30'} - \nabla \Psi_{41'} -3 \nabla \Psi_{50'}),$$
$$(\widetilde{{\rm curl}}~ Q)_{13}  \cong 
{5\sqrt{2}\ii \over {4}} (\nabla \Psi_{10'} +\nabla \Psi_{21'}+
\nabla \Psi_{30'} - \nabla \Psi_{41'}),$$
and 10 similar complex (20 more real) equations.

\subsection{Identifying the matter derivatives in $\nabla R$}

The $16$ matter variables in ($1+3$) form (given by the $8\ 
\{\widetilde{\nabla}_a\mu,\,\dot{\mu},\, \widetilde{\nabla}_a p,\,
\dot{p}\}$; $6\ \{\sigma_{ab},\ \Theta\},\ 3\ \dot{u}_c$ and $3\ 
\omega_a,$ making 20 of which $4$ can be eliminated by the contracted
Bianchi identities (\ref{B1}) and (\ref{B2})), which give the
non-locally free derivatives in $\nabla R$ for the perfect fluid case, will
now be shown to be fully equivalent to the $16$ free quantities in the
spinorial treatment, (ii) $\nabla \Phi$, (iii) $\nabla \Lambda$ and
the matter form of (iv) $\Xi$. In the general case, the
36 real terms in these spinor components are all independent, but the
restriction to a perfect fluid introduces linear dependencies between
the components.

In the quantities (iv) in Section 3, $\Xi_{CABC'}$ (either side of
the Bianchi identities (\ref{spB1})), there are, in general $8$
complex ($16$ real) independent components. For our case, when these are
calculated as the right hand side of (\ref{spB1}), i.e.\ from the matter
terms, these will be functions of spin-coefficients times $\Phi_{00}$ and
partial derivatives of $\Phi_{00}$ along the null basis vectors.
We obtain, for instance,
\begin{eqnarray}
\Xi_{10'} & = & \smfrac{1}{3} (\Phi_{00,l} - \Phi_{00,k}) + \smfrac{1}{3} 
[\bar{\widehat{\mu}} - 2 \widehat{\mu} - \widehat{\rho} + 2
\bar{\widehat{\rho}} - 2 (\gamma + \bar{\gamma})]
\Phi_{00}, \nonumber \\
\Xi_{11'} & = & \smfrac{1}{3} \Phi_{00,m} + \smfrac{1}{3} 
[2 \bar{\pi} - 2 \kappa - \bar{\nu} + \tau ] \Phi_{00}, \label{Xi}
\end{eqnarray}
where the usual $\rho$ and $\mu$ of the Newman-Penrose formalism have
been written $\widehat{\rho}$ and $\widehat{\mu}$, to avoid confusion
with matter variables.

For the perfect fluid case, as we would expect from inspection of the
matter terms in (\ref{B3})-(\ref{B6}), the number of algebraically
independent quantities in $\Xi$ drops to $11$ (real) components. In
detail, we find that $\Xi_{10'}- \Xi_{21'}$ is real,
$\Xi_{01'}=\overline{\Xi_{30'}}$ and
$3(\Xi_{11'}-\overline{\Xi_{20'}}) = \Xi_{00'}- \overline{\Xi_{31'}}$
(5 real equations).

Through the ONT-null tetrad transformation, (\ref{ONT2NP}), the fluid
kinematic quantities can be expressed in terms of spin-coefficients.
Direct calculation shows that, taking $(\mu+p)$ as known from the
Riemann tensor itself, the free quantities in $\Xi$ can be taken as
the $5\ \sigma_{ab}$, the $3\ \omega_a$ and the three combinations
$(\mu+p)_{,\alpha} + \dot{u}_\alpha(\mu+p)$. Since $\Xi$ contains only
$(\mu+p)$ it cannot yield the right side of (\ref{B4}) directly. In
terms of the Weyl tensor, choosing the free quantities in $\Xi$ is
equivalent to specifying the $5$ time derivatives of $E_{ab},\ 
\dot{E}_{ab},$ and $6$ divergences of $E_{ab}$ and $H_{ab}$, $(
\widetilde{{\rm div}} E )_a,\ ( \widetilde{{\rm div}} H )_a,$ taking
matter to be free.

As examples of these relations, we give
\begin{eqnarray}
6 \sqrt{2}\left( \Xi_{10'} + \overline{{\Xi}_{10'}} \right) & = &
-2 ((\mu+p)_{,3} + \dot{u}_3 (\mu + p)) - 3 \sigma_{33} (\mu + p) 
\nonumber \\
\sqrt{2} i \left( \Xi_{10'} - \overline{{\Xi}_{10'}} \right) & 
= & -\omega_{12} (\mu + p) = -\omega_{3} (\mu + p), \nonumber \\
\sqrt{2} i \left( \Xi_{01'} - \overline{{\Xi}_{01'}} \right) & 
= & -\sigma_{12} (\mu + p). \label{rel}
\end{eqnarray}

In the general case, there are also $16$ (real) algebraically
independent quantities in the (ii) in Section 3, i.e.\
${\nabla_{(X}}^{(X'} {\Phi_{AB)}}^{A'B')}$ (the totally symmetrized spinor
derivative of the Ricci spinor).  In our case,  these
quantities take the form
\begin{eqnarray}
\nabla \Phi_{00'} & = & \Phi_{00,k} - 2 (\varepsilon + \bar{\varepsilon})
\Phi_{00}\nonumber\\
\nabla \Phi_{01'} & = & \smfrac{1}{3} \Phi_{00,m} 
+ \smfrac{2}{ 3} ( - \bar{\alpha}
- \beta - \bar{\pi} + \kappa) \Phi_{00}\label{DPhi}\\
\ldots & &\nonumber
\end{eqnarray}
One component ($\nabla \Phi_{03'}$) is zero, while the others obey the
5 real identities given by $\nabla \Phi_{02'}=\nabla \Phi_{13'}$,
$3(\nabla \Phi_{01'}+ \nabla \Phi_{23'})= \nabla \Phi_{12'}$ and
$\nabla \Phi_{00'}+9\nabla \Phi_{22'}=9\nabla \Phi_{11'}+\nabla
\Phi_{33'}$, so that for a perfect fluid are only 9 independent real
components within $\nabla \Phi$.  By direct calculation, we find that
these 9 independent quantities can be given by the $5\ \sigma_{ab}$,
3 expressions $(\mu+p)_{,\alpha} - 2\dot{u}_\alpha(\mu+p)$, and $(\mu+p)\dot{}
-2\Theta(\mu+p)/3$.

For example, we find
\begin{eqnarray}
4 \sqrt{2} (\nabla \Phi)_{00'} & = & 
(\mu+p)_{,3} - 2 \dot{u}_3 (\mu + p) + (\mu+p)\dot{} - (\smfrac{2}{ 3} \Theta
+2 \sigma_{33}) (\mu + p),
\nonumber \\
12 \sqrt{2} (\nabla \Phi)_{01'} & = & 
(\mu+p)_{,1} - 2 \dot{u}_1 (\mu + p) 
 - 4 \sigma_{31} (\mu + p) \nonumber \\
& & -i \, 
[ (\mu+p)_{,2} - 2 \dot{u}_2 (\mu + p) - 4 \sigma_{23} 
(\mu + p)] .\nonumber 
\end{eqnarray}

The 5 independent real relations between components of $\Xi$ and
$\nabla \Phi$ implied by the fact that each gives the 5 $\sigma_{ab}$
can be expressed as $3\nabla \Phi_{02'}=2\Xi_{01'}$; $3(\nabla
\Phi_{23'}- \nabla \Phi_{01'})= 6\Xi_{11'}-2\Xi_{00'}$; and 
$\Xi_{10'}+\overline{\Xi_{10'}} - (\Xi_{21'}+\overline{\Xi_{21'}}) =
6\nabla\Phi_{11'} -3\nabla\Phi_{22'} - \nabla\Phi_{00'}$.  Thus $(\mu+p)$,
$\Xi$ and $\nabla \Phi$ together give the following 15 fluid
derivative quantities: the 5 $\sigma_{ab}$, 3 $\omega_a$, 3 $\dot{u}_a$,
3 $(\mu+p)_{,\alpha}$ and $(\mu+p)\dot{} -2\Theta(\mu+p)/3$. [Of
course, where we take the independent quantities to occur and which
are then regarded as dependent is not uniquely defined.]


For a perfect fluid, the partial derivatives of $\Lambda,$ (iii) in
Section 3, are related to the partial derivatives of $\Phi_{00'}$, via
the contracted Bianchi identities (\ref{spB2}) which can be written
\begin{eqnarray}
\Phi_{00',l} - \Phi_{00',k} & = & 6 ( \Lambda_{,l} - \Lambda_{,k} )
+ 4 [(\gamma + \bar{\gamma}) + (\varepsilon + \bar{\varepsilon}) ] 
\Phi_{00'},\nonumber \\ 
\label{Bi}
\Phi_{00',m} & = & 6 \Lambda_{,m} + 2 [ -\bar{\nu} + \kappa - \bar{\pi}
+ \tau ] \Phi_{00'}, \\
\Phi_{00',l} + \Phi_{00',k} & = & -2 ( \Lambda_{,l} + \Lambda_{,k} )
+ \smfrac{4}{ 3} [(\gamma + \bar{\gamma}) - (\varepsilon + \bar{\varepsilon}) 
- (\widehat{\mu} + \bar{\widehat{\mu}}) + (\widehat{\rho} +
\bar{\widehat{\rho}})] \Phi_{00'} \nonumber. 
\end{eqnarray}
In general the four derivatives of $\Lambda$ are mutually
independent. 

For a perfect fluid
\begin{eqnarray}
4 \sqrt{2} (\Lambda_{,m} + \Lambda_{,\bar{m}}) & = & 
\smfrac{1}{3} \mu_{,1} - p_{,1}, \nonumber \\
4 \sqrt{2} i (\Lambda_{,m} - \Lambda_{,\bar{m}}) & = & 
\smfrac{1}{3} \mu_{,2} - p_{,2}, \nonumber \\
4 \sqrt{2} (\Lambda_{,k} - \Lambda_{,l}) & = & 
\smfrac{1}{3} \mu_{,3} - p_{,3}. \nonumber
\end{eqnarray}
so the Bianchi identities enable us to separate the spatial
derivatives of $\mu$ and $p$. We have one remaining free
quantity,
$$12 \sqrt{2} (\Lambda_{,k} + \Lambda_{,l}) = \dot{\mu} - 3 \dot{p},$$
which together with the remaining Bianchi identity (the last in
(\ref{Bi})) and $(\mu+p)\dot{} -2\Theta(\mu+p)/3$ enables us to obtain
$\dot{p}$, $\dot{\mu}$ and $\Theta$ separately. We now have all the
fluid quantities in spinor form.

To summarize, we have shown that for the perfect fluid case the set of
quantities (ii)--(iv), i.e.\ $\nabla \Phi$, $\nabla
\Lambda$  and $\Xi,$ contains
$16$ algebraically independent real quantities (as functions of
spin-coefficients times $\Phi_{00'}$ and partial derivatives of
$\Phi_{00'}$ in the basis $1$-forms), which is the same number as
calculated for the data characterizing the non-locally free part of
$\nabla R$ in the $1 + 3$ formalism, and the two sets are equivalent.


\section{Summary}

By decomposing the gradient of the curvature $\nabla R$ in the perfect
fluid case we can identify the $24$ Maartens {\it et al.\/}\@ \cite{MES}
locally free data (curls and {\it distortions} of the electric and
magnetic part of the Weyl tensor) with the $\nabla \Psi$ quantities in
the approach to the equivalence problem \cite{MA}. The remaining
quantities in the first derivatives of the curvature, which can be
given by the $16$ algebraically independent matter variables given by
the derivatives of $\mu$ and $p,\ \Theta,\ \sigma_{ab},\ \dot{u}_c$
and $\omega_a,$ constrained by the contracted Bianchi identities, were
explicitly shown to be fully equivalent to the remaining $\Xi,\ \nabla
\Phi$ and $\nabla \Lambda$ in the spinor treatment \cite{MA}. Together
these give a full ONT - NP identification of the $24 + 16 = 40$ free
quantities characterizing $\nabla R,$ for the perfect fluid case.


\end{document}